\documentclass[
]{ceurart}

\usepackage{listings}
\usepackage{twemojis} % image-based emojis; no font changes
\usepackage[most]{tcolorbox} % for nice boxed content
\begin{document}

\copyrightyear{2025}
\copyrightclause{%Copyright for this paper by its authors.
National Research Council Canada. 
  Use permitted under Creative Commons License Attribution 4.0
  International (CC BY 4.0).}

\conference{Identity-Aware AI workshop at 28th European Conference on Artificial Intelligence,
  October 25, 2025, Bologna, Italy}

\title{From Perceived Effectiveness to Measured Impact: Identity-Aware Evaluation of Automated Counter-Stereotypes}

\author[1]{Svetlana Kiritchenko}[%
orcid=0000-0003-2550-3918,
email=svetlana.kiritchenko@nrc-cnrc.gc.ca,
url=https://svkir.com/,
]
\cormark[1]
\address[1]{National Research Council Canada, Ottawa, Canada}

\author[2]{Anna Kerkhof}[%
email=anna.kerkhof@gmx.de,
]
\address[2]{ifo Institute for Economic Research, Germany}

\author[1]{Isar Nejadgholi}[%
orcid=0000-0001-6241-6114,
email=Isar.Nejadgholi@nrc-cnrc.gc.ca,
]

\author[1]{Kathleen C. Fraser}[%
orcid=0000-0002-0752-6705,
email=kathleen.fraser@uottawa.ca,
]

\cortext[1]{Corresponding author.}

\begin{abstract}
We investigate the effect of automatically generated counter-stereotypes on gender bias held by users of various demographics on social media. Building on recent NLP advancements and social psychology literature, we evaluate two counter-stereotype strategies -- counter-facts and broadening universals (i.e., stating that anyone can have a trait regardless of
group membership) -- which have been identified as the most potentially effective in previous studies. We assess the real-world impact of these strategies on mitigating gender bias across user demographics (gender and age), through the Implicit Association Test and the self-reported measures of explicit bias and perceived utility. 
Our findings reveal that actual effectiveness does not align with perceived effectiveness, and the former is a nuanced and sometimes divergent phenomenon across demographic groups. While overall bias reduction was limited, certain groups (e.g., older, male participants) exhibited measurable improvements in implicit bias in response to some interventions. Conversely, younger participants, especially women, showed increasing bias in response to the same interventions. These results highlight the complex and identity-sensitive nature of stereotype mitigation and call for dynamic and context-aware evaluation and mitigation strategies. 
\end{abstract}

\begin{keywords}
Gender stereotypes \sep
counter-stereotypes \sep
real-world impact assessment 
\end{keywords}

\maketitle

\section{Introduction}

Despite advances over the past decades, important hurdles remain on the path to gender equality. In particular, gender\footnote{In this work, we focus on stereotypes about women and consider gender bias as the association of men and women with traditional masculine and feminine roles. This is a limited view of gender bias, and more work is needed to address other types of gender bias, including bias against non-binary people.} stereotypes persist \citep{bertrand2020gender}. Gender stereotypes reflect general expectations about the attributes, characteristics, and roles of different genders. For example, assertiveness and dominance are often ascribed to men, warmth and care for others are attributed to women \citep{fiske2010venus}. Recent empirical evidence demonstrates that gender stereotypes affect how we perceive others and 
ourselves, confining both personal choices and professional careers \citep{ellemers2018gender}. Thus, addressing gender stereotypes is critical. 

While gender stereotypes have always been ubiquitous in our society, their prevalence on social media is a relatively new phenomenon \citep{fraser2022computational,kerkhof2023gender}. A growing literature develops novel Natural Language Processing 
(NLP) techniques to measure the extent to which gender stereotypes (and other types of toxic language) exist on social media \citep{charlesworth2021gender,castano2021internet,asr2021gender,lei2024systematic}, but the analysis of potential counter-measures has received less attention. 

Trying to censor stereotypical content from online communications may be infeasible and even undesirable, as this poses a threat to the right to the freedom of speech. Influencing the users and trying to change their inner beliefs can provide a more effective and lasting solution, improving our offline interactions as well. 
One promising avenue to address stereotype propagation online is responding with counter-statements, a.k.a. counter-stereotypes. 
Counter-stereotypes challenge gender (and other types of) stereotypes; e.g., a counter-stereotype might present factual arguments against the gender stereotype, provide counter-examples, or state that a specific trait is not unique to a particular gender. 
Frequent exposure to stereotypes makes the stereotypical associations stronger, while the presence of counter-statements may potentially weaken these associations \citep{kawakami2000just}. It has previously been shown that while changing the view of the original speaker can be challenging, counter-stereotypes can have a large positive impact on the online community \citep{mivskolci2020countering,hsueh2015leave}. 

While manually crafting counter-statements can be costly, given the large volume of online communications, they can be generated automatically with Large Language Models (LLMs) and other state-of-the-art computational methods. Recent work in NLP has proposed and evaluated viable generation techniques \citep{allaway2022towards, fraser2023makes, mun2023beyond, nejadgholi2024challenging}, but their actual effect on users' beliefs has not been investigated. 
In the present work, we fill this gap and examine the question of how effective automatically generated counter-stereotypes are in challenging gender stereotypes held by users of social media. Further, we investigate whether the impact of counter-stereotypes varies across user demographics.

We conduct an online experiment (between-subject design) to assess the real-world impact of statements countering stereotypes about women on social media users. Specifically, we build on the findings of \citet{nejadgholi2024challenging} and assess counter-stereotypes generated with ChatGPT. In their online study, two counter-stereotype strategies were identified as the most potentially effective, so we focus on these two strategies in our study:  
\begin{itemize}
    \item \textbf{Broadening Universals:} Stating that the stereotypical trait is not unique to the target group and that all people, regardless of group membership, can have the trait.
    \item \textbf{Counterfacts:} Providing facts that contradict the stereotype.
\end{itemize}
The main goal of our experiment is to examine whether and how effectively these counter-stereotypes can reduce the \textit{implicit} gender bias held by users of different demographics, specifically at the intersection of age and gender. 
The implicit gender bias is measured through the Implicit Association Test (IAT) as the strength of association of women with family and home-related attributes and men with careers \citep{greenwald1998measuring}.

In an online study with more than 1200 participants, we present users with stereotypical and counter-stereotypical social-media style statements and evaluate their impact through a series of questions designed to measure (1) implicit gender bias, (2) explicit gender bias, and (3)  
perceived utility of counter-stereotypes. 
Our results demonstrate that reducing implicit (as well as explicit) gender bias on social media is a challenging task. Diverse and extended strategies might be required to positively affect users from different demographic groups. In particular, younger users, and especially younger women, might benefit from more nuanced ways of addressing stereotypes about women. 
Moreover, we found that the strategies perceived by users as likely to be effective might not actually reduce their biases.  
Further work on effectively addressing gender bias in online communications is needed.

\section{Related Work}

Our research draws on previous work in social psychology and natural language processing. 
In the following section, we briefly summarize some of the related work on measuring and countering stereotypical beliefs.

\subsection{Psychological Studies on Countering Stereotypes}

Numerous psychological studies focused on stereotypes, prejudices, and biased attitudes. 
However, directly assessing people's biases through self-reported measures of attitudes may be insufficient since study participants may choose not to share their actual preferences due to social desirability bias or may not even be aware of their automatic biased associations. 
Thus, most psychological assessments complement self-reported \textit{explicit} measures with indirect or \textit{implicit} measures of attitudes. Bias measured with explicit measures is referred to as \textit{explicit bias}, and bias measured with implicit measures is referred to as \textit{implicit bias} \citep{greenwald2020implicit}. One frequently used implicit measure is the Implicit Association Test \citep{greenwald1998measuring}.  IAT has been designed as a procedure to indirectly measure the strength of association of concepts with attributes (e.g., the concept of race with positive or negative sentiment, or the concept of gender with science or art) in human participants. 
Since its introduction, the IAT has been validated and widely applied to a variety of concepts and attributes \citep{greenwald2020implicit}.

Through the use of implicit measures, previous studies demonstrated that 
stereotypes and biased associations are malleable given appropriate strategies and conditions \citep{blair2002malleability,forscher2019meta}. A number of such strategies have been proposed, including asking participants to consider others' perspectives, broadening the association to members of other groups, or inducing empathy and positive emotions \citep{fraser2023makes}. 
For example, \citet{dasgupta2001malleability} showed pictures of admired Black and disliked White individuals to participants and noticed reduced implicit racial bias as measured with the IAT. In another experiment, participants who were asked to imagine a counter-example (a strong woman) also demonstrated lower implicit gender bias (measured with IAT) as compared to participants engaged in stereotypic (a weak woman), neutral (a vacation in the Caribbean), or no imagery task \citep{blair2001imagining}. 
\citet{lai2014reducing} used IAT along with self-report measures of racial attitudes to compare 17 interventions for reducing racial preferences and found that interventions featuring exposure to counter-stereotypical exemplars, evaluative conditioning (e.g., showing Black faces with positive words), and intentional strategies to overcome bias (e.g., setting an intention to respond positively to a Black face) were the most effective in reducing implicit bias. 
Such strategies present information incongruent with the original stereotype and may help weakening stereotypical associations \citep{kawakami2000just}. 

Still, uncertainties remain on the effectiveness of implicit bias reduction techniques \citep{forscher2019meta,fitzgerald2019interventions}. In the experiment by \citet{morin2017counter}, fourth-grade girls actually performed \textit{better} on a math test when the title page of the test had a stereotypical image (an icon of a boy with a comic-like speech bubble saying, ``I'm very good at geometry''), while boys showed higher performance when a counter-stereotypical image (an icon of a girl with the same speech bubble) was present. A study by \citet{rudman2010effect} 
showed that exposure to non-traditional roles (e.g., a female surgeon and a male
nurse) through biography reading decreased women's leadership self-concept and resulted in lower interest in traditionally masculine occupations. Further, \citet{palffy2023countering} found that counter-stereotypical framing and role modeling were effective in increasing the number of applications from young women to STEM jobs, but did not help in increasing the number of applications from young men to stereotypically female jobs in healthcare.

In the present work, we test two counter-stereotype strategies, counterfacts and broadening universals, and assess whether these strategies can effectively reduce implicit gender bias in the reader in a simulated social media environment. We also examine whether the effect varies for different demographic groups. 

\subsection{Identifying and Tackling Stereotypes with NLP}

Much of the work on stereotypes in NLP has focused on detecting biased associations in language models and other NLP tools \cite{bolukbasi2016man,caliskan2017semantics, may2019measuring,liang2021towards}.
Other relevant work has developed NLP methods for detecting stereotypes in \textit{human-written} text \cite{cryan2020detecting,fraser2022computational, bosco2023detecting, liu2024quantifying,cignarella2025stereotype}. 
In contrast, our work investigates the question of how to address stereotypical writing online in a useful way.

Various automatic and semi-automatic ways of generating counter-statements have been proposed in the literature \citep{qian-etal-2019-benchmark,mathew2019thou,chung-etal-2021-towards,tekiroglu-etal-2020-generating}. More recently, large language models have been successfully employed for this task. However, without specific instructions on counter-stereotype strategies, LLMs tend to output generic statements simply denouncing the stereotypes \citep{mun2023beyond}. 

Several studies examined specific counter-stereotype strategies in combination with the LLM use.  
\citet{allaway2022towards} explored several different methods for automatically countering ``essentialism'', or the belief that members of a group are somehow essentially alike. Essentialist beliefs underlie many stereotypes. In an online annotation study, participants ranked the strategies of \textit{broadening universals} (defined above), and \textit{tolerance} (a generic counter-statement reminding the reader that we should be tolerant of others' differences) as the ``most effective''.
\citet{mun2023beyond} also conducted a study of annotator preferences for six different counter-stereotype strategies. When presented with a stereotype and human-written examples of each strategy, annotators consistently ranked \textit{broadening universals} as the ``most convincing'', followed by \textit{alternate qualities} (a statement emphasizing alternative qualities of group members), and \textit{counter-examples}. 

\citet{fraser2023makes} identified 11 possible counter-stereotype strategies and used GPT-3.5 to generate examples from each category. A set of four annotators labeled the examples for perceived quality. 
Overall, annotators preferred \textit{warning of consequences}, \textit{showing empathy}, and \textit{denouncing} stereotypical statements, though there were differences depending on the nature of the stereotype (e.g., descriptive versus prescriptive). 
\citet{nejadgholi2024challenging} considered the same 11 strategies in a study in which crowd-workers were asked to rate ChatGPT-generated statements countering common gender stereotypes. The counter-statements were evaluated for their offensiveness, plausibility, and potential effectiveness. While the obtained ratings showed different trends for male and female annotators, the two strategies, \textit{broadening universals} and \textit{counterfacts}, emerged as the most robust approaches for both groups. 

In all of the studies discussed above, counter-stereotypes were evaluated for \textit{perceived} quality, by asking annotators which methods they preferred or believed to be effective. 
In contrast, \citet{bar2024generative} examined the \textit{real-world effect} of automatically generated counter-statements on the authors of  hateful posts on X and found that LLM-generated statements promoting empathy or warning of consequences were largely ineffective in changing the speakers' attitudes. 
In the current study, we also assess the actual impact of counter-stereotypes, but focus on their effect on the \textit{reader's} attitudes as well as whether and how such effect varies across demographics.

\section{Study Design}

We conducted our study on the effect of counter-stereotypes on gender bias in two phases. First, we ran a small-scale pilot study, followed by the full-scale study. In both pilot and full-scale studies, the set-up was the same. We will first give an overview of the study design, and then go into detail on each component. 

\subsection{Study Overview}

The study was conducted using the Prolific crowd-sourcing platform\footnote{\url{https://www.prolific.com/}} and programmed using oTree \citep{chen2016otree}. We recruited participants who were at least 18 years old, fluent in English, and resided in the United States. We also requested a gender-balanced sample. 

Participants were randomly assigned to one of the four treatment groups:

\begin{itemize}
    \item \textbf{Control:} Participants were presented with 10 short social media style conversations (statement--response) on gender-neutral topics (travel, nature, pets, etc.), e.g., 

\begin{tcolorbox}[colback=white, colframe=black, boxrule=0.5pt, sharp corners, left=4pt, right=4pt, top=4pt, bottom=4pt]
\footnotesize 

\noindent $\triangleright$ Statement: ``Embracing the great outdoors, one trail at a time.
\texttwemoji{1f332}\ \texttwemoji{26f0} \#HikingAdventures \#NatureLover \#ExploreTheWild''

\medskip

\noindent $\triangleright$ Response: ``Absolutely love the thrill of hiking! It's like a journey
to the soul's happy place. \texttwemoji{1f97e}\ \texttwemoji{1f33f}
\#HikingIsLife \#NatureTherapy \#AdventureTime''

\end{tcolorbox}

    \item \textbf{Stereotype:} Participants were presented with 5 conversations on neutral topics, as above, and 5 conversations containing a gender stereotype paired with a neutral response, e.g.,

\begin{tcolorbox}[colback=white, colframe=black, boxrule=0.5pt, sharp corners, left=4pt, right=4pt, top=4pt, bottom=4pt]
\footnotesize 

\noindent $\triangleright$ Statement: "C'mon, women are just bad at science. 
\texttwemoji{1f468-1f3fd}\ \texttwemoji{1f469-1f3fd} \#BoyzNGirlz"

\medskip

\noindent $\triangleright$ Response: "Taking in all sides of the conversation before forming an opinion. 
\texttwemoji{1f914} \#ListeningMode"

\end{tcolorbox}

    \item \textbf{Broadening Universals:} Participants were presented with 5 conversations on neutral topics, and 5 conversations containing a gender stereotype paired with a counter-statement using the Broadening Universals strategy, e.g.,

\begin{tcolorbox}[colback=white, colframe=black, boxrule=0.5pt, sharp corners, left=4pt, right=4pt, top=4pt, bottom=4pt]
\footnotesize 

\noindent $\triangleright$ Statement: ``Well, women do not drive well. \texttwemoji{1f697}\ \texttwemoji{1f469-1f3fd} \#truthbetold \#lifelessons''

\medskip

\noindent $\triangleright$ Response: ``C'mon... Driving ability is not determined by gender and is unique to each individual, let's leave stereotypes in the past. \texttwemoji{1f697}\ \texttwemoji{1f469-1f3fd}\ \texttwemoji{1f468-1f3fd} \#EqualityInDriving''

\end{tcolorbox}

    \item \textbf{Counterfacts:} Participants were presented with 5 conversations on neutral topics, and 5 conversations containing a gender stereotype paired with a counter-statement using the Counter-facts strategy, e.g., 

\begin{tcolorbox}[colback=white, colframe=black, boxrule=0.5pt, sharp corners, left=4pt, right=4pt, top=4pt, bottom=4pt]
\footnotesize 

\noindent $\triangleright$ Statement: ``If you ask me, women are dependent. \#realtalk \#factsoflife''

\medskip

\noindent $\triangleright$ Response: ``Simple fact: women make up 47\% of the US labor force and are primary or co-breadwinners in 63\% of households, proving they are not dependent. \texttwemoji{1f469-1f3fd}''
\end{tcolorbox}

\end{itemize}

All participants were first asked to consent to participate in the study. If they agreed, they were provided with a brief description and the instructions for the survey. 
Next, they were shown the 10 social media style conversations as described above (one conversation per screen), according to their treatment group assignment. After that, an online version of the Implicit Association Test (IAT)\footnote{The IAT interface was similar to \url{https://implicit.harvard.edu/implicit/takeatest.html}} was conducted. 
Following that, we assessed the participants' judgments of the strategies directly by asking them to rate how much they enjoyed reading the responses and whether they thought those responses were effective in challenging gender stereotypes. As a second measure of counter-stereotype utility assessment, we employed an incentive-compatible Becker-deGroot-Marschak mechanism (BDM), widely used in experimental economics to measure the utility of an item (a product, an outcome, etc.) for an individual. 
Then, we measured the participants' \textit{explicit} gender bias through a set of five questions. 
Finally, participants were asked to self-report their gender (male, female, or other) and their age.

The entire study took on average 17 minutes to complete. Each participant received \$2.50 USD (plus a potential BDM payment) upon completion of the full survey. The study was approved by the Research Ethics Boards of the authors' institutions.

\subsection{Conversation Generation}

\paragraph{Control Conversations:} Ten conversations, with one statement and one response each, were automatically generated using ChatGPT (gpt-3.5-turbo). The prompts provided the gender-neutral topic of the conversation (e.g., travel, nature, pets) and instructed ChatGPT to generate short, tweet-style texts.

\paragraph{Stereotypes and Counter-stereotypes:} 
We used five negative, descriptive gender stereotypes against women and the corresponding counter-stereotypes automatically generated by ChatGPT from the study by \citet{nejadgholi2024challenging}. 
The stereotypes were selected such that the automatically generated counter-stereotypes for the two selected strategies were labeled by crowd-workers in \citep{nejadgholi2024challenging} as both non-offensive and plausible.\footnote{The actual factuality of the counter-statements was not assessed.} The crowd-workers also rated the selected counter-stereotypes as potentially effective, with the average rating of 0.35 on a -1 to 1 scale for the counter-facts and 0.13 for the Broadening Universals counter-statements. 
The neutral responses (e.g., ``\textit{Exploring the pros and cons of this issue, eager to hear diverse viewpoints. \#OpenMind}'',  ``\textit{Interesting topic, looking forward to learning more. \#CuriousMinds}'') for the Stereotype treatment group were also generated automatically with ChatGPT. The stereotype statements and counter-statements were manually adjusted to make the conversations sound more natural (e.g., by adding phrases like 'actually', 'if you ask me', etc.) Also, related hashtags were appended to the end of the stereotype statements.

The five stereotypes are the following:

\begin{tcolorbox}[colback=white, colframe=white, boxrule=0.5pt, sharp corners,
  left=4pt, right=4pt, top=4pt, bottom=4pt]
\footnotesize 

\begin{enumerate}
\item Simple fact: women are emotionally unstable. \#JustAFact \#SoTrue

\item C'mon, women are just bad at science. \texttwemoji{1f468-1f3fd} \texttwemoji{1f469-1f3fd} \#BoyzNGirlz

\item Honestly, women are not very assertive. \#justsaying

\item Well, women do not drive well. \texttwemoji{1f697} \texttwemoji{1f469-1f3fd} \#truthbetold \#lifelessons

\item If you ask me, women are dependent. \#realtalk \#factsoflife
\end{enumerate}

\end{tcolorbox}

Broadening Universals counter-statements:

\begin{tcolorbox}[colback=white, colframe=white, boxrule=0.5pt, sharp corners,
  left=4pt, right=4pt, top=4pt, bottom=4pt]
\footnotesize 

\begin{enumerate}
\item Geez.. It's unfair to label women as emotionally unstable when humans of all genders can display emotional instability. \texttwemoji{1f469-1f3fd} \texttwemoji{1f468-1f3fd} \#EndStereotyping

\item Really, let's stop perpetuating the harmful stereotype that all individuals who struggle with science are inherently bad at it, regardless of gender or other identity factors. \#STEMEquality

\item Honestly, assertiveness is not exclusive to a particular gender or group, it is a personality trait that varies among individuals. \#BreakTheStereotype

\item C'mon... Driving ability is not determined by gender and is unique to each individual, let's leave stereotypes in the past. \texttwemoji{1f697} \texttwemoji{1f469-1f3fd} \texttwemoji{1f468-1f3fd} \#EqualityInDriving

\item Now, really ... relying on others for assistance and support is a human trait, not exclusive to one gender. \texttwemoji{1f469-1f3fd} \texttwemoji{1f468-1f3fd} \#DependenceIsNotGenderSpecific
\end{enumerate}

\end{tcolorbox}

Counterfacts:

\begin{tcolorbox}[colback=white, colframe=white, boxrule=0.5pt, sharp corners,
  left=4pt, right=4pt, top=4pt, bottom=4pt]
\footnotesize 

\begin{enumerate}
\item You know, studies have found no significant difference in emotional stability between men and women. \texttwemoji{1f645-200d-2642-fe0f} \texttwemoji{1f645-200d-2640-fe0f} \#StopTheStereotype

\item Actually, women earn approximately 50\% of all science and engineering bachelor's degrees and make up 45\% of the life sciences workforce. \#WomenInSTEM \texttwemoji{1f469-1f3fd}

\item C'mon... Women are just as assertive as men, as research shows that there is no significant difference in levels of assertiveness between genders. \texttwemoji{1f645-200d-2642-fe0f} \texttwemoji{1f645-200d-2640-fe0f} \#GenderEquality

\item Look, studies show that women are just as safe and competent drivers as men, with fewer accidents and traffic violations documented compared to men. \texttwemoji{1f697} \texttwemoji{1f469-1f3fd} \#WomenCanDriveWell

\item Simple fact: women make up 47\% of the US labor force and are primary or co-breadwinners in 63\% of households, proving they are not dependent. \texttwemoji{1f469-1f3fd}
\end{enumerate}

\end{tcolorbox}

\subsection{Implicit Association Test (IAT)}

In this study, we used the version of IAT that measures the association of binary genders (male, female) with career or family \citep{xu2018gender}, which aligns best with the stereotypes chosen for the study. The gender concept was represented through common male and female first names. The attribute words were work-related (e.g., business, office, salary) or family-related (e.g., children, parents, wedding). The full list of concept and attribute words is provided in Table~\ref{tab:stimuli}.

We followed the standard procedure consisting of seven blocks of trials \citep{greenwald2003understanding}. In each block, participants needed to categorize words representing a concept or an attribute as quickly as possible by pressing one of the two prespecified keys on a keyboard. In the first two blocks, words representing the concepts and the attributes were categorized separately. Then, the stimulus words for both concepts and attributes were categorized simultaneously, first with female-representing and family-related words corresponding to one key and male-representing and career-related words with the other key (stereotypical set-up), and second with female-representing and career-related words corresponding to one key and male-representing and family-related words with the other key (anti-stereotypical set-up). For participants who stronger associate females with domestic, family-centered environment, the second set-up should be harder and result in longer response times. The IAT score is mainly determined by the difference in response times between the two set-ups.   

We used 14 stimuli in blocks 1, 2, and 5 (where words representing the concepts and the attributes were presented separately to the participants), and 28 stimuli in the main blocks 3, 4, 6, and 7 (where concepts and attributes were categorized simultaneously). For scoring, we applied the algorithm by \citet{greenwald2003understanding} as follows. We used the response times in the main blocks 3, 4, 6, and 7, and discarded trials with response times greater than 10,000 ms. We applied a penalty mechanism to deal with categorization errors: for each participant, the response times for incorrect responses were replaced with the mean response times of correct responses in the same block plus a penalty of 600 ms. Then, two separate scores were calculated for blocks 3 and 6, and blocks 4 and 7, as the difference between the mean response times in the two blocks divided by the pooled standard deviation in the two blocks. The final D-score was obtained as the average of these two scores. Higher D-scores indicate stronger association of females with family and males with career (implicit stereotype).

\begin{table}[t]
\caption{Stimuli for the Implicit Association Test \citep{xu2018gender}.}
\centering
\begin{tabular}{p{8cm}} 
\toprule
\textbf{Gender concept words}\\
$\ \ \ \ $\textbf{Female:} Anna, Emily, Gina, Julia, Rebecca, Sarah, Susan\\
$\ \ \ \ $\textbf{Male:} Arthur, Ben, Daniel, Jeffrey, John, Paul, Timothy\\[5pt]
\textbf{Attribute words}\\
$\ \ \ $\textbf{Career:} business, career, corporation, management, office, professional, salary\\
$\ \ \ \ \ $\textbf{Family:} children, family, home, marriage, parents, relatives, wedding\\
\bottomrule
\end{tabular}
\label{tab:stimuli}
\end{table}

\subsection{Measuring Explicit Gender Bias}

We asked how strongly participants agree with each of the following five statements that convey common gender stereotypes \citep{swim1995sexism}:
\begin{enumerate}
    \item Women are generally not as smart as men.
    \item I would be equally comfortable having a woman as a boss than a man.
    \item It is more important to encourage boys than to encourage girls to participate in athletics.
    \item Women are just as capable of thinking logically as men.
    \item When both parents are employed and their child gets sick at school, the school should call the mother rather than the father.
\end{enumerate}
Participants could indicate their agreement with each statement on a visual analog scale from 0 to 100. We converted the answers to questions 2 and 4 by subtracting them from 100, to align all the results in one direction: higher reported values indicated stronger agreement with stereotypical gender roles. 
Then, an average score over the five questions was calculated and divided by 100. 

Participants' answers to these direct questions may be affected by \textit{social-desirability bias} -- the tendency of survey respondents to answer questions in a manner they believe is socially appropriate. In particular, the participants in our study may under-report their stereotypical beliefs. To partially mitigate this effect, we added a second set of questions about explicit gender stereotypes, which allowed the participants to report their tendencies in a more covert, anonymous manner. We asked ``With how many statements would you agree?'' for two lists of statements about personal preferences.

First list:
\begin{itemize}
    \item I prefer Indian food over Italian food.
    \item My favorite color is blue.
    \item Historical novels are boring.
    \item Beer is better than wine. 
    \item ***Men are better leaders than women.
\end{itemize}

Second list:
\begin{itemize}
    \item Football is more fun than swimming.
    \item ***Women are better caretakers than men.
    \item Apples taste better than bananas.
    \item Mathematics in school is very difficult.
    \item I prefer sunny weather over rain.
\end{itemize}

Statements marked with *** were shown only to half of the participants (randomly chosen). Since the responses do not directly reveal the individual's opinion on gender preferences, participants may be more open to provide answers aligned with their actual beliefs. However, aggregated over a \textit{group} of respondents, answers to these questions can reveal explicit bias of the group. If we assume that for a large enough random sample of participants the average number of agreements to four non-gendered statements should remain approximately the same, the difference between the average answer for the participants shown five statements (four non-gendered + the stereotype) and the participants shown four (non-gendered) statements would indicate the explicit bias of the group. The obtained score can be interpreted as a proportion of participants that agreed with the stereotypical statement.

\subsection{Assessing Perceived Utility of Counter-Stereotypes}

First, we asked participants in the two counter-strategy groups to indicate how strongly they agree with each of the following statements:
\begin{itemize}
    \item I \textit{enjoyed} reading the responses to the stereotypes.
    \item I think those responses are \textit{effective} in challenging gender stereotypes.
\end{itemize}
They provided their answers by moving a slider on a visual analog scale from 0 to 100.

We also measured the participants' level of interest (or perceived utility) in encountering stereotypical and counter-stereotypical messages on social media platforms. 
There would be a stronger motivation for social media platforms to implement measures such as automated counter-stereotyping if they perceived it as increasing user satisfaction and engagement. To assess if users \textit{value} the presence of counter-stereotypes, we employ an incentive compatible Becker-DeGroot-Marschak mechanism \citep{becker1964measuring} that elicits participants' ``willingness to accept'' (WTA) reading further statements that are similar to those that they have already seen. Intuitively, if a task is unpleasant, then people will demand more money to agree to do the task again. Participants could enter the minimal monetary reward, $B$, that they would want to be paid for reading ten further statements that resemble those that they have already seen. 
In this way, $B$ represents the participant's evaluation of the difficulty or unpleasantness of the task in the form of the required monetary reward. Then, a random amount $P$ between \$0 and \$2.00 was drawn automatically. If $P$ was higher than or equal to the participant's minimal compensation request $B$, the participant was shown ten more conversations and was paid the extra bonus $P$. If $P$ was lower than $B$, no further conversations were shown, and no bonus was paid. 
Thus, to increase their chances of getting the bonus payment $P$, the participants would need to enter the \textit{minimal} amount of compensation for this extra task. 
We winsorized the BDM bids at the upper end at 99\% (i.e., the values above the 99th percentile were set to the 99th percentile) to reduce the impact of outliers. 

\subsection{Data Verification} 

To ensure the quality of responses, we employed the following strategies:

\begin{itemize}
    \item We included a check question after the instructions to verify that the participants read the instructions carefully, and excluded participants who answered incorrectly.
    \item We added three more questions after the ten conversations, asking the participants to recall the topics of the conversations they had just read. We excluded participants who did not answer at least two of the three questions correctly. 
    \item We monitored response times in IAT and excluded participants whose response times were unrealistically short. 
\end{itemize}

All participants who completed the study were paid, even if their data was discarded.

\section{Results}

There were 1,402 unique Prolific users who attempted the main study. Of these, 1,296 completed the full survey. Six participants failed the attention question in the instructions, and 20 participants answered incorrectly two or more questions about the topics of the conversations they read. The results for these participants were not included in the following analysis. We also excluded the results for 23 participants whose responses in the IAT were too quick (more than 10\% of responses took less than 300 ms), as recommended by \citet{greenwald2003understanding}. There remained 1,247 participants (621 female, 606 male, 20 other; age: 40 $\pm$ 13). In the pilot study, out of 210 unique Prolific users who attempted the study, 198 completed the full survey. Two participants failed the attention question, five answered incorrectly two or more topic questions, and another seven participants were excluded due to high rate of quick (less than 300 ms) IAT responses. There remained 184 participants (91 female, 91 male, 2 other; age: 39 $\pm$ 14).

\subsection{Implicit Bias}

\begin{figure}[t]
\centering
\includegraphics[width=4cm]{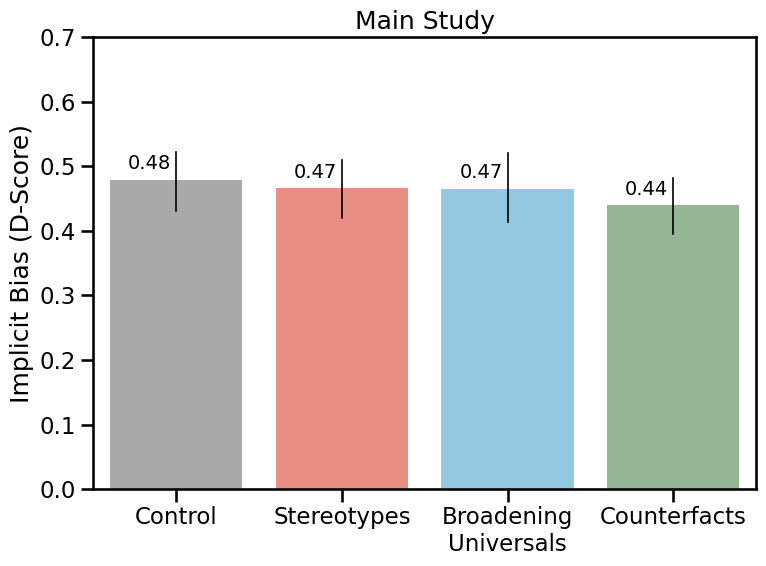}
\includegraphics[width=4cm]{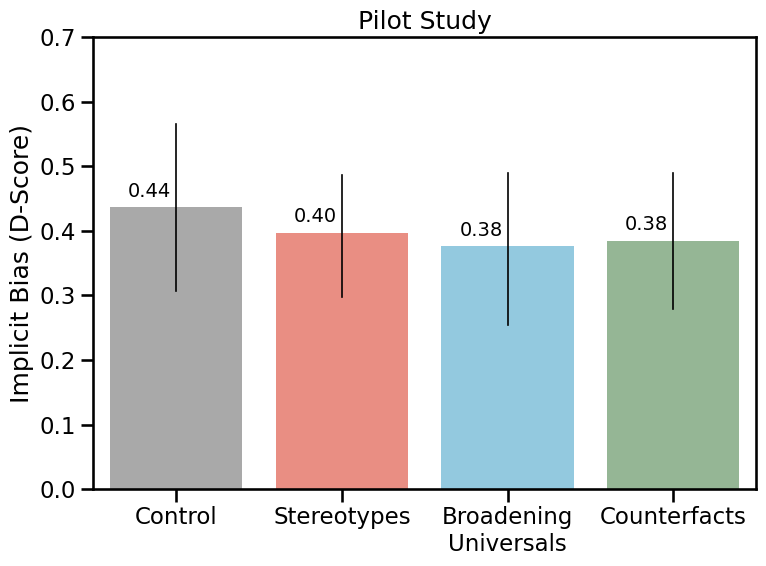}
\caption{The mean D-scores for the four treatment groups in the main and the pilot studies. Error bars indicate 95\% confidence intervals.}
\label{fig:D-treatment}
\end{figure}

The mean D-scores for the four treatment groups are presented in Figure~\ref{fig:D-treatment} (Main Study). Recall that lower D-scores imply a smaller timing discrepancy between stereotypical and anti-stereotypical blocks in the IAT, and therefore lower bias.  One can observe that both strategies, broadening universals and counterfacts, result in similar or slightly lower D-scores than the control and stereotype groups. Yet, the differences are very small and are not statistically significant.\footnote{For all results in this section, statistical significance is measured using Welch's t-test \citep{welch1947generalization}.} However, the results of the pilot study (Figure~\ref{fig:D-treatment}, Pilot Study) show similar trends. We hypothesize that the small observed differences are due to the small scale of intervention (only five stereotypical statements and counter-stereotypical responses were shown to the participants). Longer and repeated exposure to these strategies might result in more pronounced differences, though it remains a question for future studies. 

\begin{figure}[t]
\centering
\includegraphics[width=4cm]{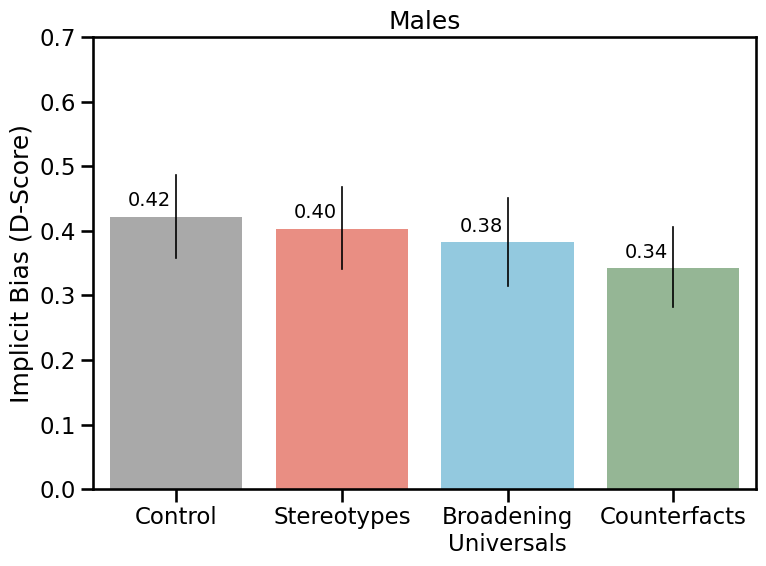}
\includegraphics[width=4cm]{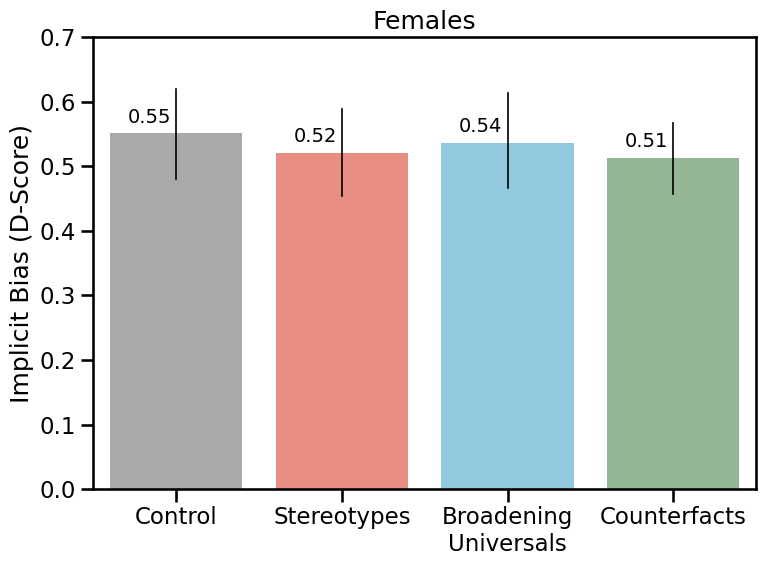}
\caption{The mean D-scores for the four treatment groups for male and female participants in the main study. Error bars indicate 95\% confidence intervals.}
\label{fig:D-gender}
\end{figure}

When looking separately at self-identified male and female participants\footnote{The group of the participants who self-identified as `other' for gender is too small for a reliable analysis.} (Figure~\ref{fig:D-gender}) in the main study, we observe that male participants in the Counterfacts condition show a bigger reduction in implicit bias relative to the other conditions, and significantly so when compared to the Control group ($p = 0.08$). 
Interestingly, female participants in the Stereotype condition demonstrated a similar or even lower implicit bias than those in the Counter-stereotype conditions, which is the opposite trend from what we had expected and from what we observed in the male participants. This may indicate that the mere presence of a stereotype against women triggers a negative response in readers, especially women, with the result equivalent to seeing the explicit counter-statement.  
We also observe that for all four treatment groups, the average implicit bias in women is higher than in men.

\begin{figure}[t]
\centering
\includegraphics[width=4cm]{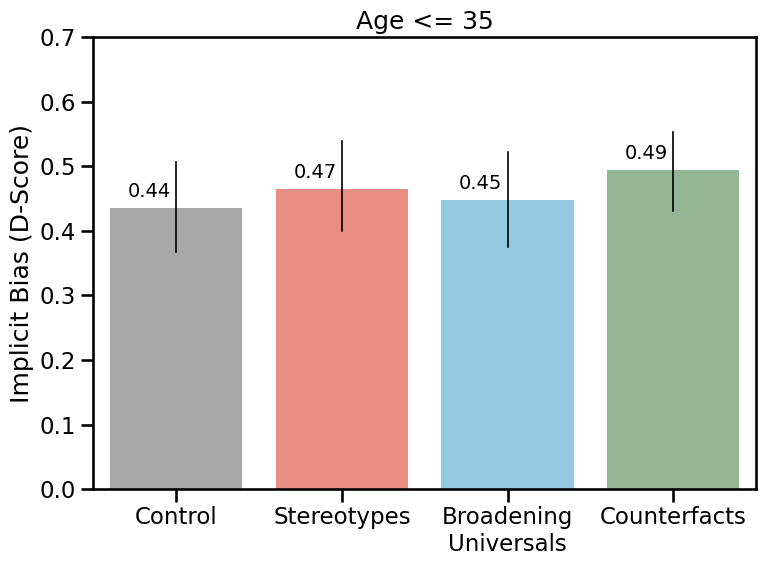}
\includegraphics[width=4cm]{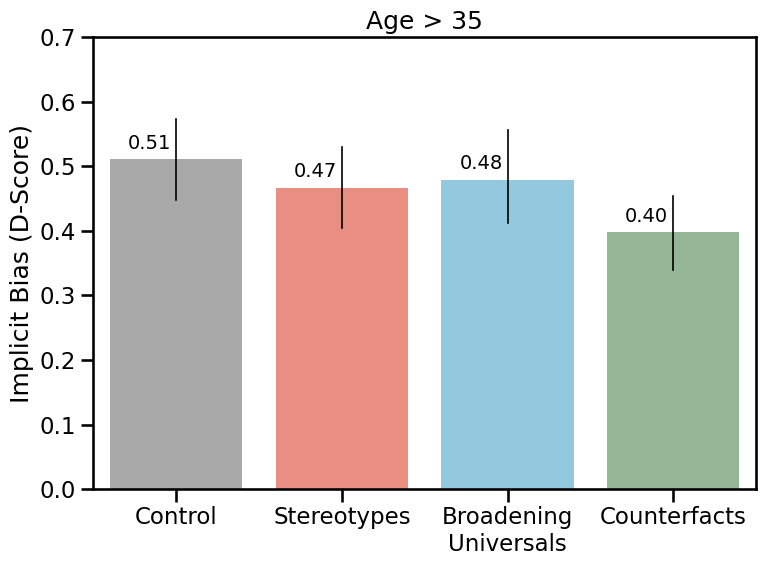}
\caption{The mean D-scores for the four treatment groups for younger ($\leq$35 y.o.) and older (>35 y.o.) participants in the main study. Error bars indicate 95\% confidence intervals.}
\label{fig:D-age}
\end{figure}

\begin{figure}[t]
\centering
\includegraphics[width=3.9cm]{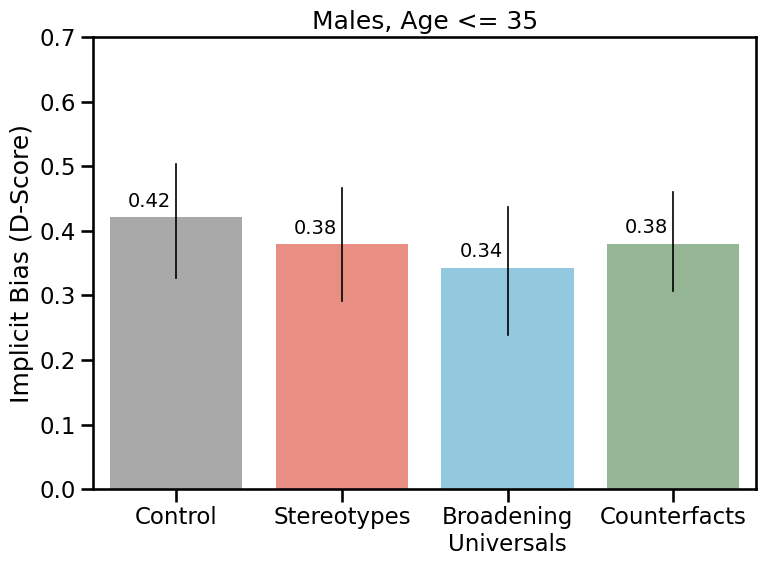}
\includegraphics[width=3.9cm]{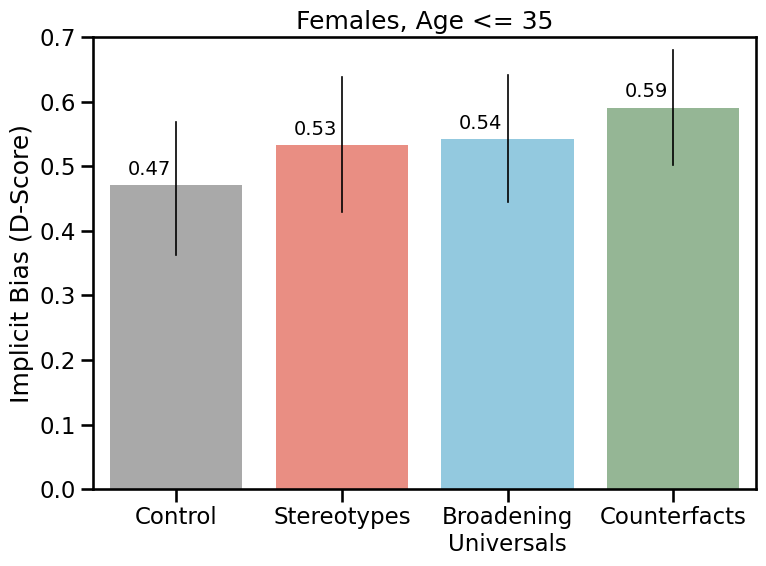}
\includegraphics[width=3.9cm]{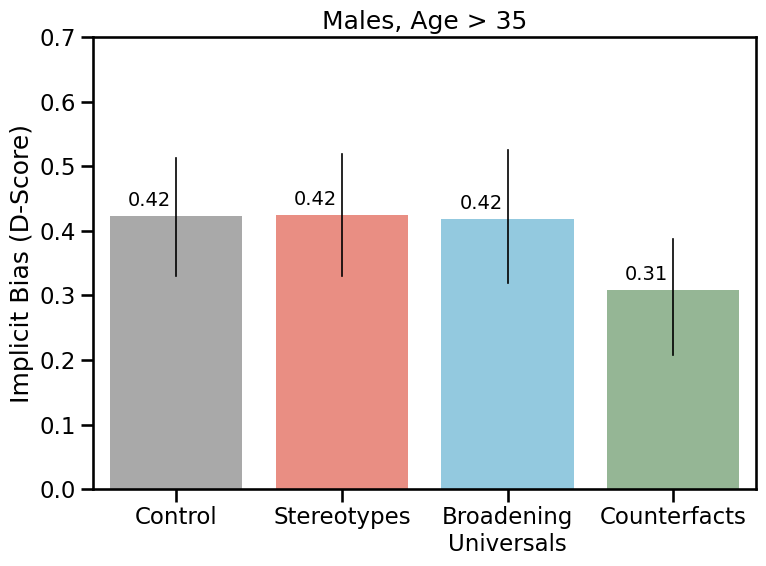}
\includegraphics[width=3.9cm]{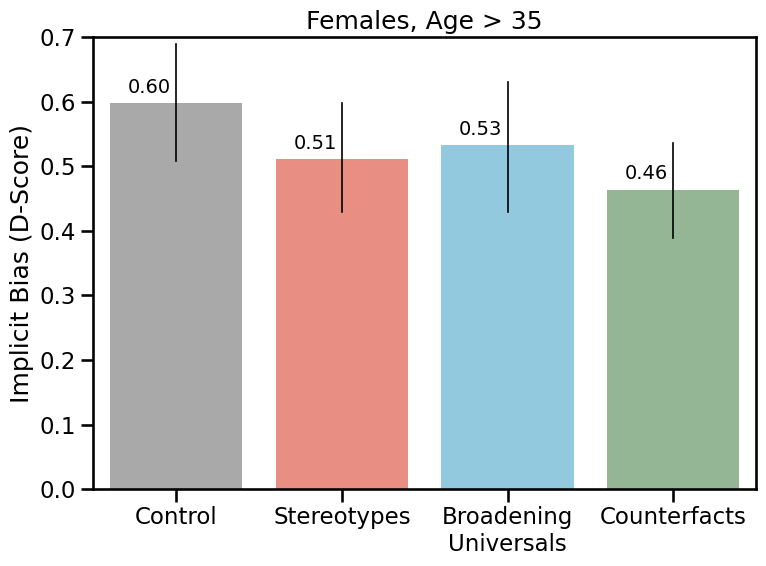}
\caption{The mean D-scores for the four treatment groups for younger ($\leq$35 y.o.) and older (>35 y.o.) male and female participants in the main study. Error bars indicate 95\% confidence intervals.}
\label{fig:D-gender-age}
\end{figure}

We also separated the participants by age into a younger group (age $\leq$ 35) and an older group (age > 35) (Figure~\ref{fig:D-age}).\footnote{The boundary of 35 y.o. was chosen as the commonly used age boundary that was close to our sample median of 37.0 y.o.} The results for these two groups are strikingly different. Older participants show reduced implicit bias after exposure to both counter-stereotype strategies relative to the Control group (the difference is statistically significant in the case of Counterfacts, $p = 0.01$). However, younger participants show the opposite pattern with the lowest implicit bias scores in the Control group. Exposure to the Counterfacts strategy actually results in the \textit{highest} implicit bias score in the younger group. The younger group also has a lower implicit bias score in the Control group than the older group (0.44 versus 0.51). Then, any presence of stereotypes, either countered or not, results in higher implicit bias. Looking intersectionally, we find this pattern is especially strong in younger female participants (Figure~\ref{fig:D-gender-age}) (the difference between the Counterfacts and Control groups is statistically significant, $p = 0.09$), while younger males as well as both (male and female) older groups show the expected pattern of decreased implicit bias after viewing counter-stereotypes as compared to the Control groups. The Counterfacts strategy appears to be effective at reducing implicit bias for both male and female participants in the older age group (the differences between the Counterfacts and Control groups are statistically significant, $p = 0.03$ and $p = 0.08$  for older female and male participants, respectively).

\subsection{Explicit Bias}

\begin{figure}[t]
\centering
\includegraphics[width=4cm]{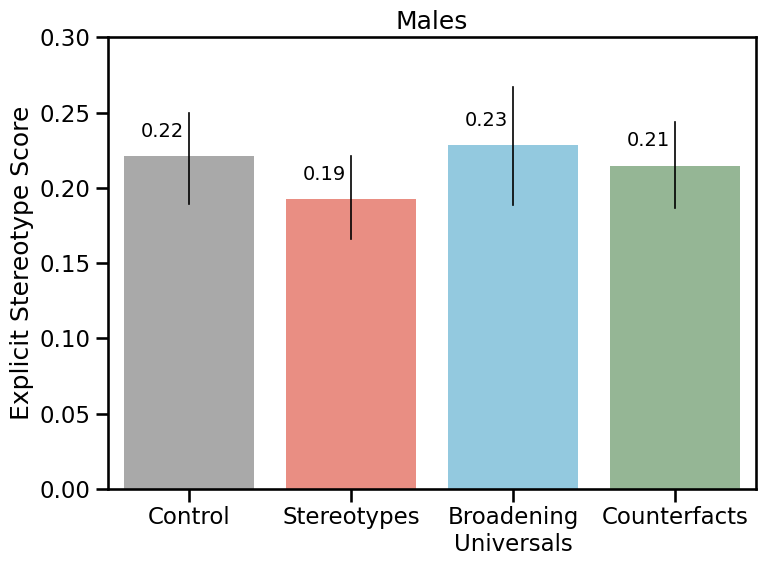}
\includegraphics[width=4cm]{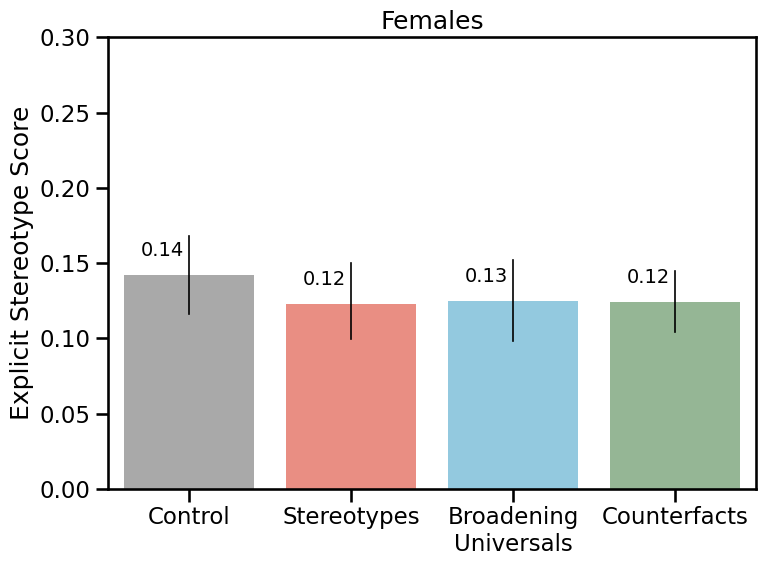}
\caption{The mean explicit stereotype scores for the four treatment groups for male and female participants in the main study. Error bars indicate 95\% confidence intervals.}
\label{fig:explicit-gender}
\end{figure}

We have also included in the survey questions that assess the self-reported explicit gender bias. Overall, participants showed low levels of explicit bias, with a mean score of 17 on a 0--100 scale of how strongly they agreed with the five statements portraying gender stereotypes. The differences between the four treatment groups are very small and not statistically significant. This is true for both male and female participants  (Figure~\ref{fig:explicit-gender}). Similarly to the implicit bias, explicit bias scores are slightly lower for the Counterfacts strategy than the Broadening Universals for both men and women. Interestingly, men exhibited substantially higher explicit bias than women in all treatment groups (Control group: 0.22 vs. 0.14, $p = 0.0002$). At the same time, as we observed earlier, the implicit bias is substantially higher in women (Control group: 0.55 vs. 0.42, $p = 0.01$). While seemingly unexpected, these findings are in line with some previous studies on gender bias \citep{salles2019estimating,kramer2021implicit}. Also, we found no correlation between explicit and implicit bias scores within the male ($r = -0.001$) or female group ($r = 0.02$).

\begin{figure}[t]
\centering
\includegraphics[width=4cm]{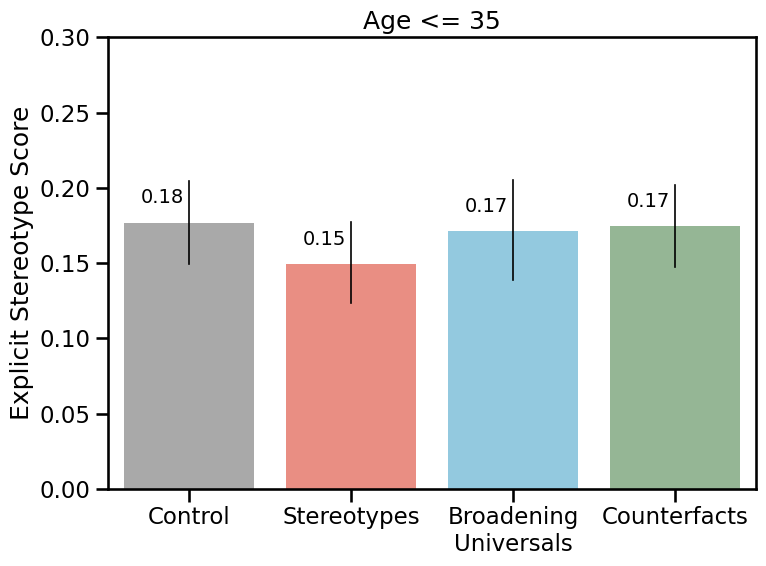}
\includegraphics[width=4cm]{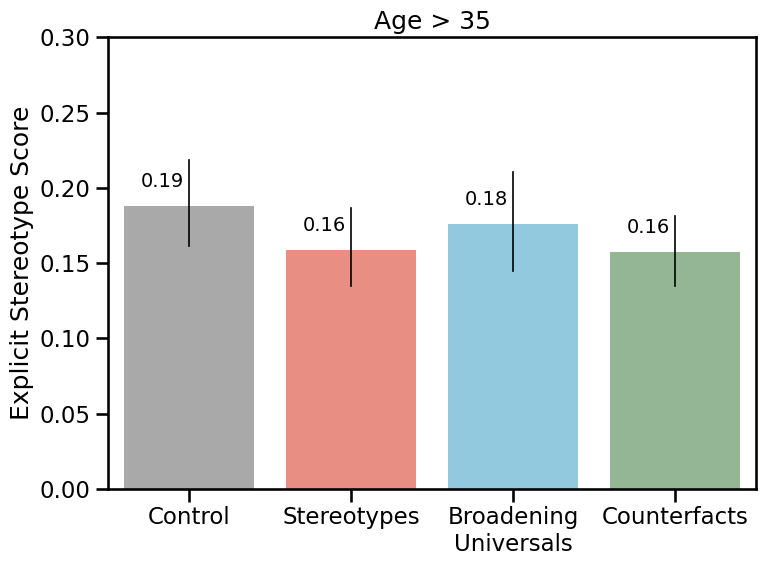}
\caption{The mean explicit stereotype scores for the four treatment groups for younger ($\leq$35 y.o.) and older (>35 y.o.) participants in the main study. Error bars indicate 95\% confidence intervals.}
\label{fig:explicit-age}
\end{figure}

For the two age groups, again the differences between the four treatment groups are very small (Figure~\ref{fig:explicit-age}). Older participants exhibited slightly lower scores in the Counterfacts group than in the Broadening Universals. 
The difference between the two age groups is negligible (e.g., for the Control group: 0.18 vs. 0.19, $p = 0.6$). However, the difference is more pronounced for the implicit bias (for the Control group: 0.44 vs. 0.51, $p = 0.12$). There is no correlation between explicit and implicit bias scores within the younger ($r = -0.1$) or older group ($r = 0.01$).  

With a more covert way of revealing the participants' explicit stereotypes (when a stereotypical statement is one of five statements and the participants report the overall number of the statements they agree with), higher levels of explicit bias are observed. About 11\% of the participants (males more frequently than females) agree that men are better leaders than women, and about 42\% of the participants agree that women are better caretakers than men. While the scores for the second question are higher for all treatment and demographic groups, the observed trends are similar between the two questions. Thus, we combine the scores for these two questions by averaging them and report the combined scores in the following analysis.

\begin{figure}[t]
\centering
\includegraphics[width=4cm]{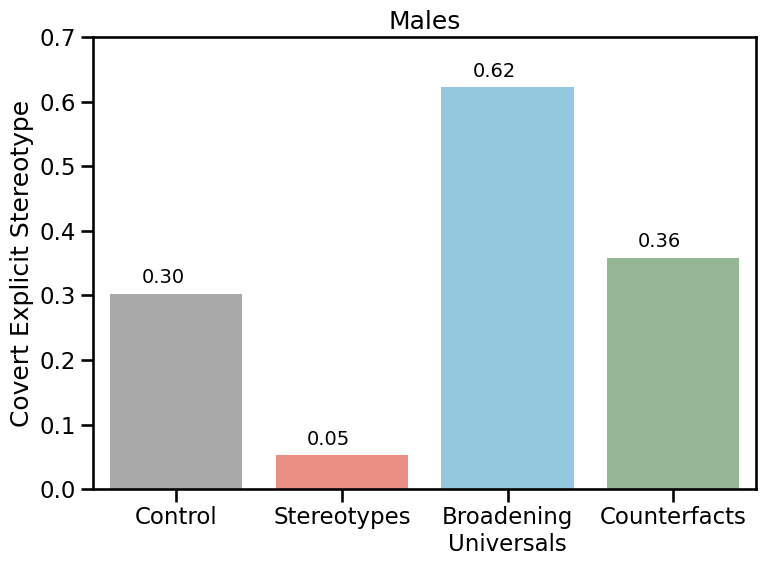}
\includegraphics[width=4cm]{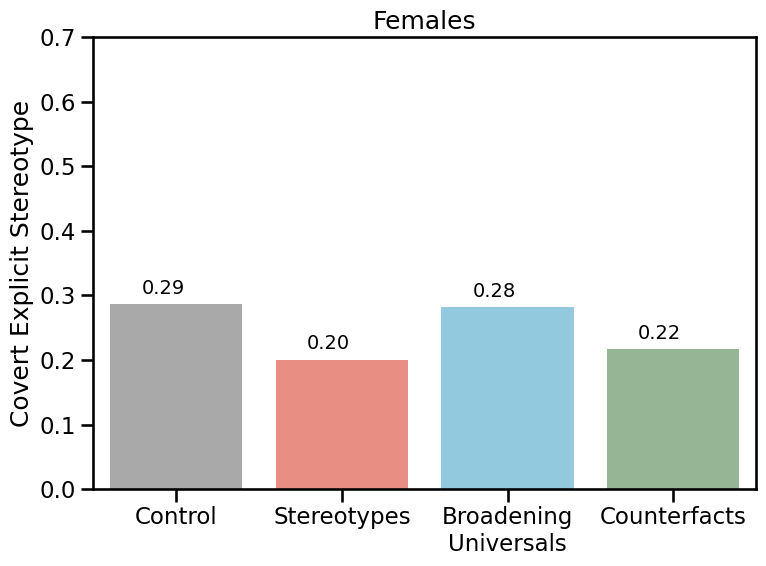}
\caption{The covert explicit stereotype scores for the four treatment groups for male and female participants in the main study.}
\label{fig:covert-gender}
\end{figure}

\begin{figure}[t]
\centering
\includegraphics[width=4cm]{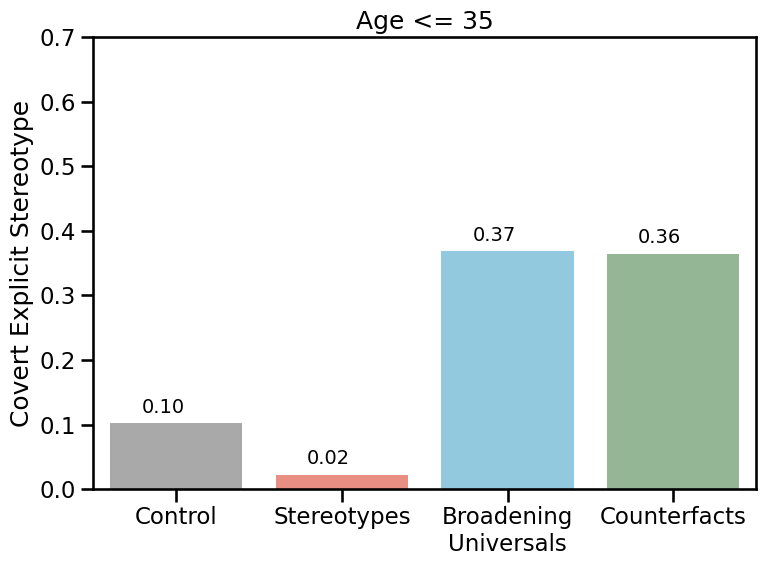}
\includegraphics[width=4cm]{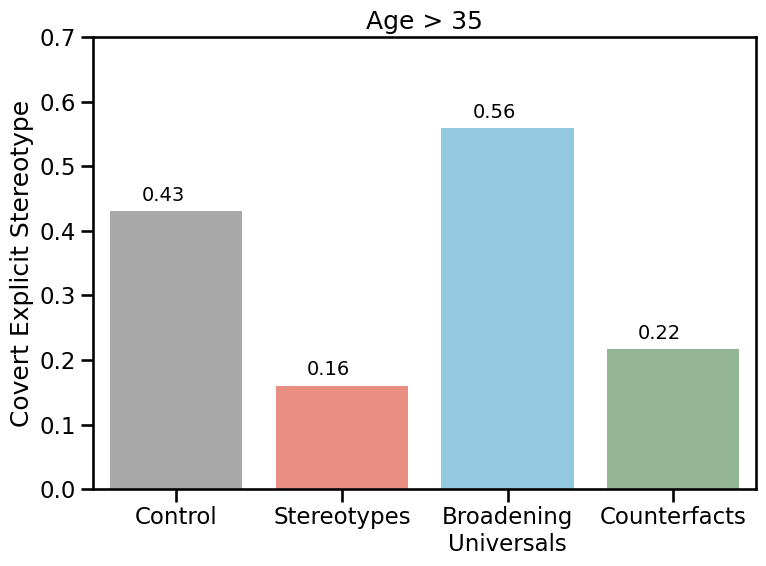}
\caption{The covert explicit stereotype scores for the four treatment groups for younger ($\leq$35 y.o.) and older (>35 y.o.) participants in the main study.}
\label{fig:covert-age}
\end{figure}

Again, we observe that men exhibit higher covert explicit bias than women (Figure~\ref{fig:covert-gender}). While differences among the four treatment groups for female participants are small, men show substantially higher scores in the Broadening Universals group. The two age groups also show different patterns (Figure~\ref{fig:covert-age}). Younger participants reveal almost no bias in the Control and Stereotypes groups, but show similarly high scores for the two counter-stereotype strategies. Older participants exhibit substantially higher bias in the Broadening Universals group than in the Counterfacts group.

Surprisingly, for all demographic groups, the lowest covert bias is shown by the Stereotypes group. The same is true for the explicit bias scores discussed above. We hypothesize that the presence of obviously stereotypical statements may trigger the sense of unfairness and result in participants trying to consciously suppress their explicit bias.

\subsection{Perceived Utility of Counter-Stereotypes}

\begin{figure}[t]
\centering
\includegraphics[width=3.9cm]{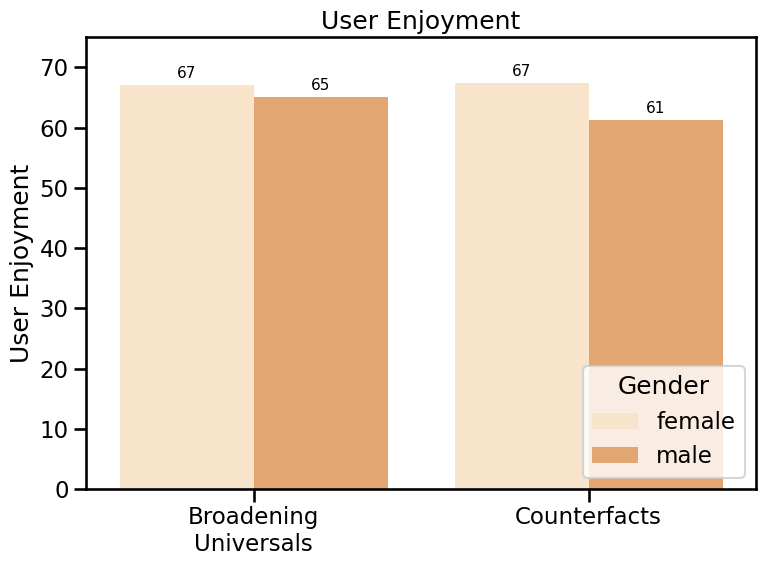}
\includegraphics[width=3.9cm]{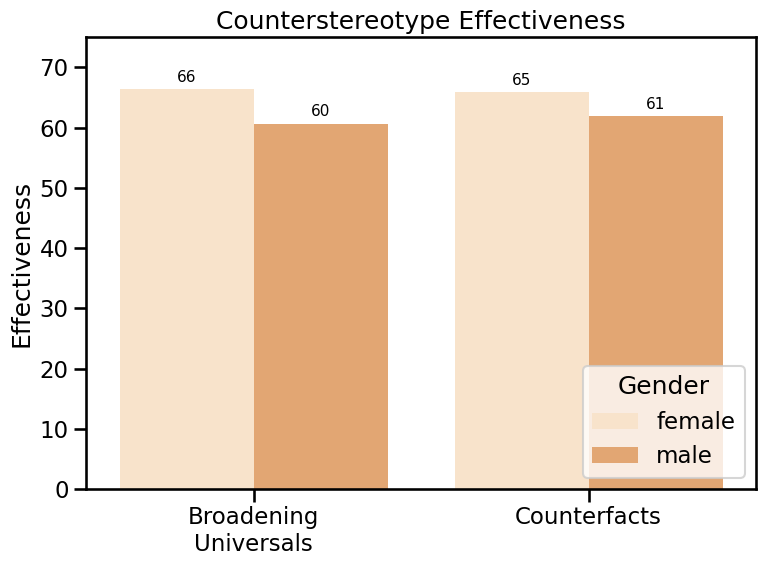}
\includegraphics[width=3.9cm]{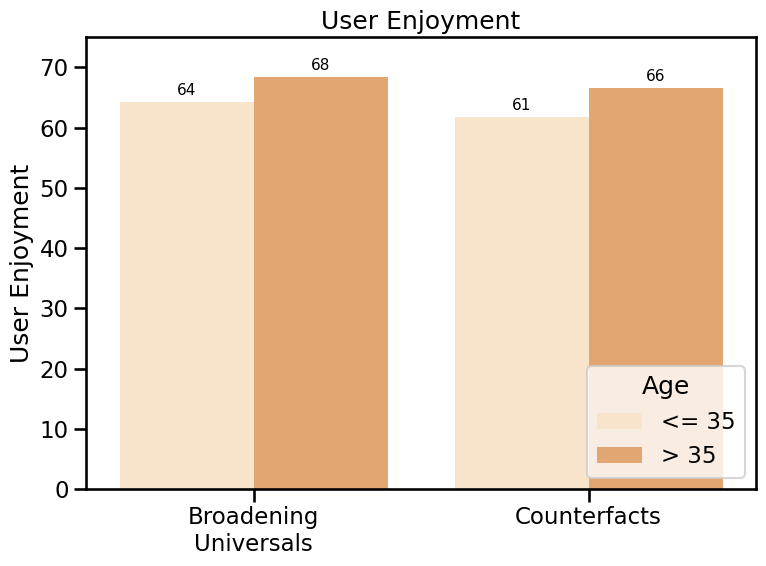}
\includegraphics[width=3.9cm]{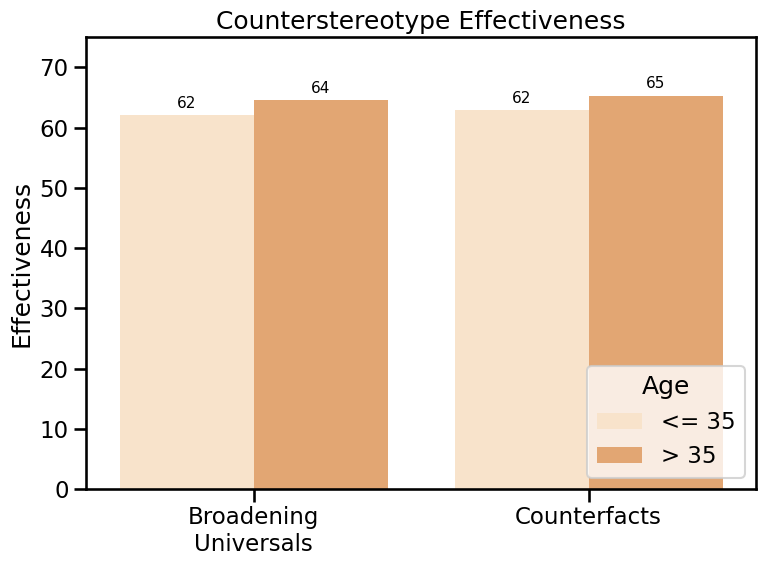}
\caption{Participants' ratings on how much they liked and how effective they found the counter-statements for the two strategies, Broadening Universals and Counterfacts.}
\label{fig:ratings-gender-age}
\end{figure}

On average, participants enjoyed reading the counter-statements, giving ratings of 66\% and 64\% to the Broadening Universals and Counterfacts, respectively. Similarly, the ratings for perceived effectiveness were 63\% and 64\%. We found moderate correlation between these ratings (Pearson $r = 0.6$). Yet again, we see a slightly different picture for different demographic groups (Figure~\ref{fig:ratings-gender-age}). Women provided higher ratings than men on enjoyment and effectiveness for both strategies. Men enjoyed Counterfacts slightly less than Broadening Universals, though rated them similarly for effectiveness. Younger participants rated both types of counter-statements lower than the older participants, giving the lowest enjoyment rating for Counterfacts.

We also found weak negative correlation between the participants' preferences and explicit bias scores: $r = -0.25$ for enjoyment and $r = -0.28$ for effectiveness. That is, participants who agreed more with gender stereotypical statements reported less enjoyment in reading counter-stereotypes, and perceived them as less effective. Conversely, there is no correlation between the participants' preferences and their implicit bias scores ($r = -0.04$ for enjoyment, $r = -0.02$ for effectiveness).

\begin{figure}[t]
\centering
\includegraphics[width=4cm]{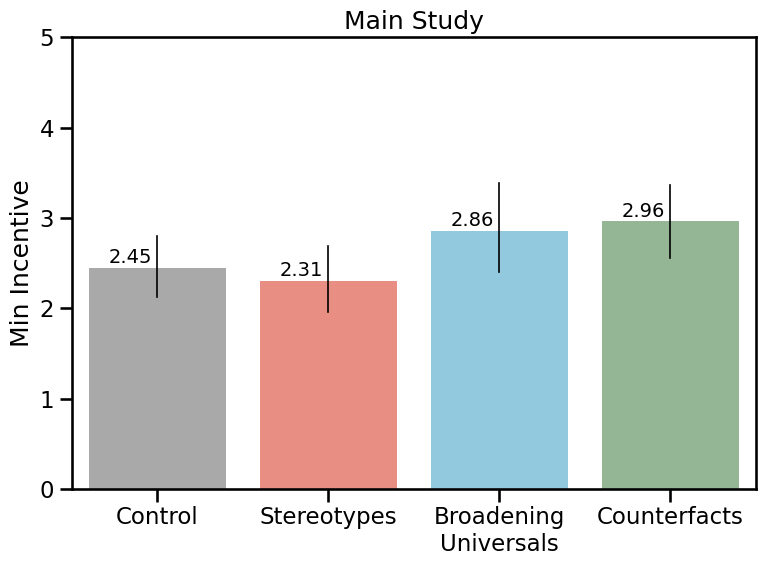}
\includegraphics[width=4cm]{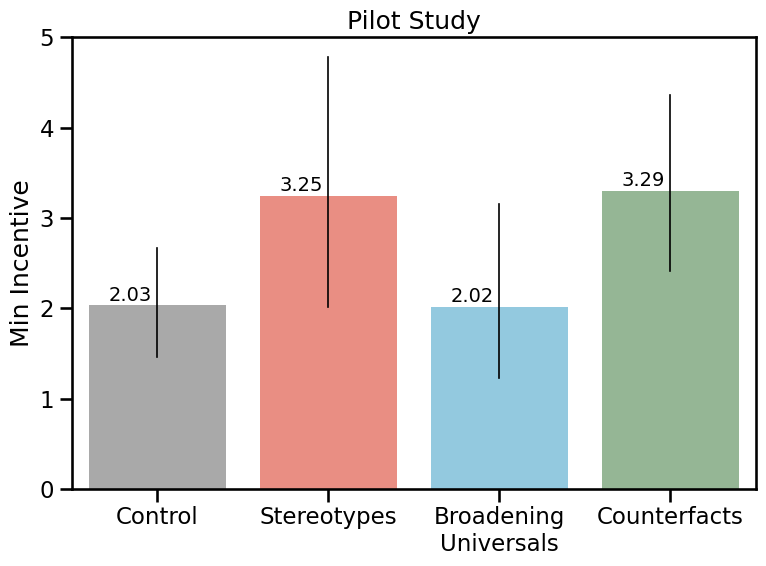}
\caption{The mean BDM bids for the four treatment groups in the main and the pilot studies. Error bars indicate 95\% confidence intervals.}
\label{fig:BDM-treatment}
\end{figure}

For the BDM experiment, Figure~\ref{fig:BDM-treatment} shows the mean BDM bids, i.e., the minimal bonus payments (or incentives) the participants were willing to accept to read more statements. The results indicate that participants were less willing to read counter-statements as compared to neutral topics or stereotypical content (their minimal bids were higher for both counter-stereotype strategies as compared to the Control and Stereotypes groups in the Main Study). 
Note, however, that the trends are quite different between the Main and Pilot studies. In the Pilot study, participants were more inclined to read further statements in the Broadening Universals group than in the Stereotype group. Also, we found no correlation between participants' enjoyment ratings and their BDM bids (Pearson $r = -0.06$). 
We hypothesize that this question may have been confusing for many participants: on one hand, participants are interested in higher bonus payments and therefore may want to put a higher bid, while on the other hand, higher bids may result in no bonus payment at all (if the random amount is lower than the bid). 
Thus, we find the results inconclusive for this experiment.

\section{Discussion}

In the current study, we did not observe significant differences in implicit gender bias between the Control, Stereotype, and two Counter-Stereotype treatment groups. The variation in participant IAT scores was substantial, and larger than the differences between the treatment groups. However, when we analyzed the results separately for various demographic groups, we found more pronounced differences. While men and women showed similar trends among the treatment groups for implicit bias, the IAT scores for women were significantly higher than the scores for men in all treatment groups. The age-based groupings showed large differences in IAT scores in the Control group, as well as strikingly different trends for the four treatment groups. Younger participants exhibited lower bias in the Control group, which increased for the Stereotype and both Counter-Stereotype treatment groups. Exposure to the Counterfacts strategy resulted in lower implicit bias than the Broadening Universals for older participants, but yielded the highest bias scores for the younger group.

Conversely, female participants exhibited substantially lower \textit{explicit} bias, both in overt and covert questions, and the two age groups showed similar levels of explicit bias. The Broadening Universals strategy resulted in higher explicit bias scores for men and older participants. In general, we observe different trends for the impact of counter-stereotype treatments on explicit and implicit bias, and no correlation between the implicit and explicit bias scores.

Interestingly, women provided higher ratings for enjoyment and effectiveness than men for both strategies, yet the counter-statements had a smaller effect on women's than on men's explicit and implicit bias. On the other hand, younger participants appreciated the counter-statements less than the older group and exhibited higher implicit, but not explicit, bias in the counter-stereotype treatment groups. 

Regarding the specific strategy used in generating the counter-stereotypes, the Counterfacts strategy tends to result in lower explicit and implicit bias scores than the Broadening Universals strategy, for all demographic groups, except younger participants. Unchallenged stereotypes (the Stereotypes group) seem to produce an unexpected effect of reducing explicit bias, but not necessarily implicit bias. 

These results suggest that countering stereotypes on social media may have different effects on various demographic groups, and even individual users. Further, explicit user preferences may not directly translate into higher reduction in either explicit or implicit user bias. 
Our measures of participants' perceived utility of counter-stereotypes found that users with higher explicit bias report liking reading the content less, and overall, participants were less willing to make competitive bids to read more counter-stereotypical content than stereotypical content. In some sense, this is not surprising: it is well-known that user engagement is driven by content that is controversial and emotional \citep{buffard2020quantitative}, and people do not like to be challenged on their beliefs \citep{kaplan2016neural}. From that perspective, social media platforms may lack the financial incentives required to engage in the generation of automated counter-stereotypes. However, allowing toxicity to flourish on a platform can lead to other financial consequences, such as loss of users and advertising revenue \citep{twitter-ad-loss}.

An encouraging finding that the younger generation tends to have lower implicit gender bias than the older participants (in the Control group) potentially indicates a positive societal shift in this area, at least in the U.S. Still, the bias exists and more work on promoting gender equality through awareness, education, and various countering strategies is needed.

\section{Conclusion}

In this study, we challenge the assumption that presenting ``seemingly effective'' counter-stereotypes reliably changes individuals' biased attitudes in real-world settings. Specifically, we showed that the impact of counter-stereotypical interventions on users' beliefs is unpredictable and not uniform across demographics. These results highlight the need for identity-aware interventions, which are dynamically tailored to audience characteristics rather than designed as one-size-fits-all solutions. 

Our study has several limitations. First, our experiments involved brief, one-off exposure to a small number of counter-stereotypical statements. This setup does not reveal the cumulative effects of repeated interventions on social media platforms. Also, we have not measured a long-term effect of these interventions on participants. While conducting a study online allowed us to reach a large pool of participants with different demographics and lived experiences, it limited our control over participants following the exact experimental procedure compared to conventional psychological studies. Our study was also limited to the US-based English-speaking participants, which leaves open questions about cross-cultural applicability. Further, other factors such as education, socio-economic status, political ideology, etc. might have an effect on gender bias, but were not examined in this study.  

Nevertheless, our results suggest that automated counter-stereotype generation has the potential to serve as a scalable strategy to mitigate stereotypical attitudes on social media. However, the interplay of demographic factors, content framing, and individual predispositions is complex and creates a tension among perceived effectiveness, actual bias reduction, and users' willingness to engage with counter-stereotypical content. We anticipate that, to make real-world impact, intervention strategies should be personalized not only in terms of content and style but also in the delivery environment, timing and potential for frequent exposure.    

Future research should explore adaptive counter-stereotyping strategies with a strong emphasis on creating measurable and sustainable attitudinal impact. Integrating multimodal delivery (text, imagery, video) and longitudinal exposure might be effective strategies to increase the impact of interventions. Cross-cultural studies are also an important direction to be pursued in future work. 

Importantly, we were only able to uncover the layered complexities of designing natural language interventions on social media because of a collaborative study design by computer scientists and behavioural economists. This interdisciplinary approach allowed us to move beyond mere perceived effectiveness measurement and observe subtle trade-offs between demographic factors, user engagement, content delivery and actual belief change. Therefore, our work highlights that research on counter-stereotyping in online discourse is not solely an NLP task, but requires sustained collaborations across computational, behavioural and social sciences to result in effective and ethically grounded interventions.

%% The declaration on generative AI comes in effect
%% https://ceur-ws.org/GenAI/Policy.html
\section*{Declaration on Generative AI}
  The authors have not employed any Generative AI tools for writing or editing the current manuscript.

%%
%% Define the bibliography file to be used
\bibliography{mybibfile}

\end{document}